\documentclass{article}%
\usepackage{amsmath}
\usepackage{amsfonts}
\usepackage{amssymb}
\usepackage{graphicx}%
\begin{document}

\title{Calculating statistical distributions from operator relations: the statistical
distributions of various intermediate statistics}
\author{Wu-Sheng Dai\thanks{Email: daiwusheng@tju.edu.cn} and Mi Xie\thanks{Email:
xiemi@tju.edu.cn}\\{\footnotesize Department of Physics, Tianjin
University, Tianjin 300072, P. R. China }\\{\footnotesize LiuHui
Center for Applied Mathematics, Nankai University \& Tianjin
University,} {\footnotesize Tianjin 300072, P. R. China}}
\date{}
\maketitle

\begin{abstract}
In this paper, we give a general discussion on the calculation of
the statistical distribution from a given operator relation of
creation, annihilation, and number operators. Our result shows
that as long as the relation between the number operator and the
creation and annihilation operators can be expressed as
$a^{\dagger}b=\Lambda\left(  N\right)  $ or $N=\Lambda^{-1}\left(
a^{\dagger}b\right)  $, where $N$, $a^{\dagger}$, and $b$ denote
the number, creation, and annihilation operators, i.e., $N$ is a
function of quadratic product of the creation and annihilation
operators, the corresponding statistical distribution is the
Gentile distribution, a statistical distribution in which the
maximum occupation number is an arbitrary integer. As examples, we
discuss the statistical distributions corresponding to various
operator relations. In particular, besides Bose-Einstein and
Fermi-Dirac cases, we discuss the statistical distributions for
various schemes of intermediate statistics, especially various
$q$-deformation schemes. Our result shows that the statistical
distributions corresponding to various $q$-deformation schemes are
various Gentile distributions with different maximum occupation
numbers which are determined by the deformation parameter $q$.
This result shows that the results given in much literature on the
$q$-deformation distribution are inaccurate or incomplete.

\end{abstract}


\section{Introduction}

The statistical property of a quantum system is embodied in the operator
relation of creation, annihilation, and number operators. When such a relation
is given, one can in principle solve the statistical distribution for the
system. For example, from $N=a^{\dagger}a$ and $\left[  a,a^{\dagger}\right]
=1$, one can deduce the Bose-Einstein distribution, and from $N=a^{\dagger}a$
and $\left\{  a,a^{\dagger}\right\}  =1$, one can deduce the Fermi-Dirac distribution.

As generalizations of Bose-Einstein and Fermi-Dirac statistics, there are some
schemes of intermediate statistics \cite{Gentile,gPauli,Wu,Wilczek}. It has
been shown that intermdiate-statistics type excitations may exist in many
physical systems \cite{Kitaev}. The need of intermediate statistics in physics
is that there are many composite-particle systems and intermediate-statistics
type elementary excitations, e.g., the Cooper pair in the theory of
superconductivity, the Fermi gas superfluid \cite{FermiSuperfluid}, the
exciton \cite{exciton}, the magnon \cite{OurJSATA}, etc. Concretely, composite
bosonic particles consisted of fermions will deviate from Bose-Einstein
statistics to a certain extent under some circumstances
\cite{Guan,CompositeParticle}. In this case, such a system can be viewed as
obeying a kind of intermediate statistics, and intermediate statistics can be
used as an effective tool for studying such a system.

Different intermediate-statistics schemes correspond to different operator
relations. In principle, from the operator relation of a kind of intermediate
statistics, one can achieve the corresponding intermediate-statistics
distribution. In the following, we will give a general discussion on the
calculation of the statistical distribution from a given operator relation.

Let $a^{\dagger}$ and $b$ be creation operator and annihilation operator, and
let $N$ be the number operator. Then we must have%
\begin{equation}
\left[  N,a^{\dagger}\right]  =a^{\dagger}\text{ \ and \ }\left[  N,b\right]
=-b. \label{NandCA}%
\end{equation}
Denoting the eigenstate of the number operator by $\left\vert N\right\rangle
$, i.e., $N\left\vert N\right\rangle =N\left\vert N\right\rangle $, we achieve%
\begin{align}
a^{\dagger}\left\vert N\right\rangle  &  =\sqrt{\alpha\left(  N+1\right)
}\left\vert N+1\right\rangle ,\nonumber\\
b\left\vert N\right\rangle  &  =\sqrt{\beta\left(  N\right)  }\left\vert
N-1\right\rangle , \label{CAactState}%
\end{align}
where the coefficients $\alpha\left(  N\right)  $ and $\beta\left(  N\right)
$ are functions of $N$. It should be emphasized that Eq. (\ref{CAactState}) is
a basic relation. From this relation, one can deduce the operator relations
among creation, annihilation, and number operators, including the quantization
condition. For example, for the Bose case, we have $b=a$, $\alpha\left(
N+1\right)  =N+1$, and $\beta\left(  N\right)  =N$; as a result, the bosonic
quantization condition reads $\left[  a,a^{\dagger}\right]  =aa^{\dagger
}-a^{\dagger}a=1$ and the relation between $a^{\dagger}$, $a$, and $N$ reads
$a^{\dagger}a=N$ or $aa^{\dagger}=N+1$. (Note that one cannot uniquely
determine the result of the creation and annihilation operators acting on a
state, i.e., the relation (\ref{CAactState}), from the quantization condition.)

Assume that the relation between the number operator and the creation and
annihilation operators is%
\begin{equation}
a^{\dagger}b=\Lambda\left(  N\right)  , \label{adbN}%
\end{equation}
where $\Lambda\left(  N\right)  $ is an analytic function, i.e., $N$ is a
function of quadratic product of the creation and annihilation operators.
Then, from Eqs. (\ref{NandCA}) and (\ref{CAactState}), we achieve%
\begin{equation}
ba^{\dagger}=\Lambda\left(  N+1\right)  . \label{badN}%
\end{equation}

The statistical distribution is the ensemble average of the number operator of
the $l$-th state%
\begin{equation}
\left\langle N_{l}\right\rangle =\frac{1}{\Xi}tr\left[  e^{-\beta\left(  H-\mu
N_{total}\right)  }N_{l}\right]  , \label{averageNl}%
\end{equation}
where $H=%
{\displaystyle\sum_{l}}
N_{l}\varepsilon_{l}$ is the Hamiltonian of the system, $N_{total}=%
{\displaystyle\sum_{l}}
N_{l}$ is the total number of particles in the system, $\mu$ is the chemical
potential, $\Xi$ is the grand partition function, $\varepsilon_{l}$ is the
energy of the $l$-th state, and $\beta=1/\left(  kT\right)  $. In the
following, we will give a general discussion on the derivation of the
statistical distribution on the basis of the operator relations among
creation, annihilation, and number operators.

As a direct generalization of Bose-Einstein and Fermi-Dirac statistics,
Gentile suggested a scheme of intermediate statistics --- Gentile statistics
in which the maximum occupation number of particles of a quantum state is a
finite number $n$ \cite{Gentile}. Bose-Einstein and Fermi-Dirac statistics
become two limit cases of Gentile statistics when $n\rightarrow\infty$ and
$n=1$, respectively \cite{OursAnn,OurDoBose}. In the following, we will show
that when the relation between the number operator and the creation and
annihilation operators takes the form of Eq. (\ref{adbN}), the statistical
distribution is always the Gentile distribution with a maximum occupation
number determined by the intermediate-statistics parameter.

As examples of the general result, we will first discuss the statistical
distributions for Bose-Einstein, Fermi-Dirac, and Gentile cases, and then
discuss the statistical distributions for various $q$-deformation schemes.

Quantum algebras (quantum groups), as generalizations of usual Lie algebras,
have been discussed widely for many years \cite{Drinfeld,Jimbo}. Quantum
algebras become important in physics since the introduction of $q$-deformed
harmonic oscillator which provides a bosonic realization of the quantum
algebra $su_{q}\left(  2\right)  $. Many schemes of $q$-oscillator have been
constructed, including the $q$-deformed Arik-Coon oscillator \cite{AC}, the
$q$-deformed Biedenharn-Macfarlane oscillator \cite{Bie}, the parabosonic and
parafermionic oscillators \cite{OK}, the $q$-deformed parabosonic and
$q$-deformed parafermionic quantization schemes \cite{OKK}, the $q$-deformed
fermionic algebra \cite{Hay}, the Tamm-Dancoff cut-off deformation \cite{OKK},
the two-parameter deformed oscillator \cite{CJ1991}, etc. Moreover, further
studies present some generalized deformed oscillators
\cite{Das,CCN,MMP,OS,Bur}. Many researches also devoted to the deformed
algebras, such as $su_{q}\left(  2\right)  $ \cite{Bie}, $su_{q}\left(
1,1\right)  $ \cite{KD,UA}, $su_{p,q}\left(  1,1\right)  $ \cite{CJ1991},
$su_{q}\left(  N\right)  $ \cite{SF}, $U_{q}\left(  gl\left(  2\right)
\right)  $ \cite{GX,KB}, $U_{p,q}\left(  gl\left(  2\right)  \right)  $
\cite{BH,CJ}, $U_{p,q}\left(  gl\left(  1,1\right)  \right)  $ \cite{BBC},
$q$-deformed Lorentz algebra \cite{SWZ,OSW,Ane}, etc.

Once the operator relation of a $q$-deformation scheme is given, the
corresponding statistical distribution is determined. In this paper, as
examples of the above general result, we discuss the statistical distributions
for various $q$-deformation schemes. We argue that the statistical
distributions of various $q$-deformation schemes are the Gentile distributions
whose maximum occupation numbers are determined by the $q$-deformation
parameter $q$ other than the distributions given in literature; note that
Bose-Einstein and Fermi-Dirac distributions are viewed as special cases of the
Gentile distribution with maximum occupation numbers $\infty$ and $1$
\cite{OursAnn}. In other words, our result indicates that the $q$-deformation
statistical distributions given in the literature are inaccurate or
incomplete. The inaccurate results in literature are obtained by an improper
approximation. Concretely, a key step in the derivation of the statistical
distribution is to deal with the average value $\left\langle f\left(
N\right)  \right\rangle $, where $f\left(  N\right)  $ is a function of $N$.
In the literature \cite{TRM}, however, such an average value is approximately
taken as $\left\langle f\left(  N\right)  \right\rangle \simeq f\left(
\left\langle N\right\rangle \right)  $, or, for more details, in the
literature the authors take the approximation $\left\langle q^{N}\right\rangle
\simeq q^{\left\langle N\right\rangle }$
\cite{Alg,SCC,SCC2,SCC3,CSC,SS,LS,AA,AA2,KLQ,LSS,Duc,LY}. Nevertheless, the
rigorous result should be $\left\langle N^{m}\right\rangle =\left\langle
N\right\rangle \left\langle N^{m-1}\right\rangle -\frac{\partial}{\partial
x}\left\langle N^{m-1}\right\rangle $ rather than $\left\langle N^{m}%
\right\rangle =\left\langle N\right\rangle ^{m}$, where $x=\beta\varepsilon$
and $\varepsilon$ is the energy of the quantum state. Moreover, in literature
there is an alternative way for considering the statistical distributions of
various $q$-deformation schemes: replacing the average of number operator
$\left\langle N\right\rangle $ by the average $\left\langle a^{\dagger
}a\right\rangle $ \cite{DK,KMD,CFM,AGI,AG,AGI2,SDA}. The reason why taking
such a replacement, as stated in the literature \cite{DK,KMD,CFM}, is that
$\left\langle N\right\rangle $ gives a nondeformed Bose-Einstein distribution.
Our result shows that the statistical distributions coming from $\left\langle
N\right\rangle $ are not only the Bose-Einstein statistics; in some cases the
statistical distribution is the Gentile distribution and the maximum
occupation number is determined by the deformation parameter $q$.

In Section \ref{deduce}, we give a general discussion on the derivation of the
statistical distribution from a given operator relation. In Section
\ref{qdeformation}, as examples, we discuss the statistical distributions for
various $q$-deformation schemes. The conclusion and outlook are given in
Section \ref{conclusion}.

\section{Deducing the statistical distribution from a given operator relation
\label{deduce}}

In this section, we give a general discussion on the calculation of
statistical distribution from the operator relation. The statistical
distribution of a physical system can be calculated from the basic operator
relations \cite{cal}. In the following, we will show that the relation between
the number operator and the creation and annihilation operators,
$\Lambda\left(  N_{l}\right)  =a_{l}^{\dagger}b_{l}$, determines the
statistical distribution. Or, in more details, the statistical distribution is
determined by the first two nonnegative integer zeroes of $\Lambda\left(
N_{l}\right)  $. In this section, we use the subscript $l$ to denote the
$l$-th state.

\subsection{The statistical distribution}

Let $p_{0}$, $p_{1}$, $p_{2}$,$\cdots$ be zeroes of $\Lambda\left(
N_{l}\right)  $. We first have the following general result:

\textit{Let }$p_{k_{1}}$\textit{\ and }$p_{k_{2}}$\textit{\ be two zeroes of
}$\Lambda\left(  N_{l}\right)  $\textit{. Then}%
\begin{equation}
\left\langle N_{l}\right\rangle =\frac{1}{z^{-1}e^{\beta\varepsilon_{l}}%
-1}-\frac{p_{k_{2}}-p_{k_{1}}}{z^{-\left(  p_{k_{2}}-p_{k_{1}}\right)
}e^{\left(  p_{k_{2}}-p_{k_{1}}\right)  \beta\varepsilon_{l}}-1}+p_{k_{1}}
\label{Nl}%
\end{equation}
\textit{is the statistical distribution corresponding to the operator relation
(\ref{adbN}), where }$z=e^{\beta\mu}$\textit{\ is the fugacity.}

The proof is as follows.

The ensemble average of $\Lambda\left(  N_{l}\right)  $ can be calculated
directly from Eq. (\ref{adbN}):%
\begin{equation}
\left\langle \Lambda\left(  N_{l}\right)  \right\rangle =\left\langle
a_{l}^{\dagger}b_{l}\right\rangle =\frac{1}{\Xi}tr\left[  e^{-\beta\left(
H-\mu N_{total}\right)  }a_{l}^{\dagger}b_{l}\right]  .
\end{equation}
By the relation%
\begin{equation}
e^{-\beta\left(  H-\mu N_{total}\right)  }a_{l}^{\dagger}=e^{-\beta\left(
\varepsilon_{l}-\mu\right)  }a_{l}^{\dagger}e^{-\beta\left(  H-\mu
N_{total}\right)  },
\end{equation}
we achieve%
\begin{align}
\left\langle \Lambda\left(  N_{l}\right)  \right\rangle  &  =e^{-\beta\left(
\varepsilon_{l}-\mu\right)  }\frac{1}{\Xi}tr\left[  e^{-\beta\left(  H-\mu
N_{total}\right)  }b_{l}a_{l}^{\dagger}\right] \nonumber\\
&  =e^{-\beta\left(  \varepsilon_{l}-\mu\right)  }\left\langle \Lambda\left(
N_{l}+1\right)  \right\rangle , \label{eqLn}%
\end{align}
\bigskip or, equivalently,%
\begin{equation}
\frac{\left\langle \Lambda\left(  N_{l}\right)  \right\rangle }{\left\langle
\Lambda\left(  N_{l}+1\right)  \right\rangle }=e^{-\beta\left(  \varepsilon
_{l}-\mu\right)  }=ze^{-x}, \label{LNbNp1}%
\end{equation}
where $x=\beta\varepsilon_{l}$.

Since $p_{k_{1}}$ and $p_{k_{2}}$ are two zeroes of $\Lambda\left(
N_{l}\right)  $, $\Lambda\left(  N_{l}\right)  $ can be expressed as%
\begin{equation}
\Lambda\left(  N_{l}\right)  =\left(  N_{l}-p_{k_{1}}\right)  \left(
N_{l}-p_{k_{2}}\right)  G\left(  N_{l}\right)  . \label{LNl}%
\end{equation}
Expanding $G\left(  N_{l}\right)  $ as%
\begin{equation}
G\left(  N_{l}\right)  =\sum_{m=0}^{\infty}c_{m}N_{l}^{m} \label{ExpGN}%
\end{equation}
and substituting Eqs. (\ref{LNl}) and (\ref{ExpGN}) into Eq. (\ref{LNbNp1})
gives%
\begin{equation}
\sum_{m=0}^{\infty}c_{m}\left\langle \left(  N_{l}-p_{k_{1}}\right)  \left(
N_{l}-p_{k_{2}}\right)  N_{l}^{m}\right\rangle =\sum_{m=0}^{\infty}%
c_{m}e^{-\beta\left(  \varepsilon_{l}-\mu\right)  }\left\langle \left(
N_{l}+1-p_{k_{1}}\right)  \left(  N_{l}+1-p_{k_{2}}\right)  \left(
N_{l}+1\right)  ^{m}\right\rangle . \label{ExpeqGN}%
\end{equation}

Now, we prove that Eq. (\ref{Nl}) is a solution of Eq. (\ref{ExpeqGN}).

First, consider the case of $m=0$.

The term with $m=0$ in Eq. (\ref{ExpeqGN}) is%
\begin{equation}%
\begin{array}
[c]{l}%
\left[  1-e^{-\beta\left(  \varepsilon_{l}-\mu\right)  }\right]  \left\langle
N_{l}^{2}\right\rangle -\left[  \left(  p_{k_{1}}+p_{k_{2}}\right)
+e^{-\beta\left(  \varepsilon_{l}-\mu\right)  }\left(  2-p_{k_{1}}-p_{k_{2}%
}\right)  \right]  \left\langle N_{l}\right\rangle \\
+\left[  p_{k_{1}}p_{k_{2}}-e^{-\beta\left(  \varepsilon_{l}-\mu\right)
}\left(  p_{k_{1}}p_{k_{2}}-p_{k_{1}}-p_{k_{2}}+1\right)  \right]  =0.
\end{array}
\label{meq0}%
\end{equation}

We first prove that%
\begin{equation}
\left\langle N_{l}f\left(  N_{l}\right)  \right\rangle =\left\langle
N_{l}\right\rangle \left\langle f\left(  N_{l}\right)  \right\rangle
-\frac{\partial}{\partial x}\left\langle f\left(  N_{l}\right)  \right\rangle
. \label{meanNfN}%
\end{equation}
The proof is straightforward:%
\begin{align}
\left\langle N_{l}f\left(  N_{l}\right)  \right\rangle  &  =\frac{1}{\Xi
}tr\left[  \frac{\partial}{\partial x}e^{-\beta\left(  H-\mu N\right)
}f\left(  N_{l}\right)  \right] \nonumber\\
&  =\left(  -\frac{1}{\Xi}\frac{\partial\Xi}{\partial x}\right)  \frac{1}{\Xi
}tr\left[  e^{-\beta\left(  H-\mu N\right)  }f\left(  N_{l}\right)  \right]
-\frac{\partial}{\partial x}\left\langle f\left(  N_{l}\right)  \right\rangle
\nonumber\\
&  =\left\langle N_{l}\right\rangle \left\langle f\left(  N_{l}\right)
\right\rangle -\frac{\partial}{\partial x}\left\langle f\left(  N_{l}\right)
\right\rangle .
\end{align}
This gives%
\begin{equation}
\left\langle N_{l}^{2}\right\rangle =\left\langle N_{l}\right\rangle
^{2}-\frac{\partial\left\langle N_{l}\right\rangle }{\partial x}, \label{N2N}%
\end{equation}
and Eq. (\ref{meq0}) is converted into a differential equation:%
\begin{equation}%
\begin{array}
[c]{l}%
\left[  1-e^{-\beta\left(  \varepsilon_{l}-\mu\right)  }\right]  \left(
\left\langle N_{l}\right\rangle ^{2}-\frac{\partial\left\langle N_{l}%
\right\rangle }{\partial x}\right)  -\left[  \left(  p_{k_{1}}+p_{k_{2}%
}\right)  +e^{-\beta\left(  \varepsilon_{l}-\mu\right)  }\left(  2-p_{k_{1}%
}-p_{k_{2}}\right)  \right]  \left\langle N_{l}\right\rangle \\
+\left[  p_{k_{1}}p_{k_{2}}-e^{-\beta\left(  \varepsilon_{l}-\mu\right)
}\left(  p_{k_{1}}p_{k_{2}}-p_{k_{1}}-p_{k_{2}}+1\right)  \right]  =0.
\end{array}
\label{dmeq0}%
\end{equation}

It can be checked directly that the statistical distribution (\ref{Nl}) is a
solution of Eq. (\ref{dmeq0}).

Next, consider the case of $m>0$.

We will prove that if%
\begin{equation}
\frac{\left\langle F\left(  N_{l}\right)  \right\rangle }{\left\langle
F\left(  N_{l}+1\right)  \right\rangle }=ze^{-x}, \label{FoF}%
\end{equation}
then%
\begin{equation}
\frac{\left\langle N_{l}^{m}F\left(  N_{l}\right)  \right\rangle
}{\left\langle \left(  N_{l}+1\right)  ^{m}F\left(  N_{l}+1\right)
\right\rangle }=ze^{-x}. \label{NFoNF}%
\end{equation}

By Eqs. (\ref{meanNfN}) and (\ref{FoF}), we have%
\begin{equation}
\left\langle \left(  N_{l}+1\right)  F\left(  N_{l}+1\right)  \right\rangle
=ze^{x}\left\langle N_{l}F\left(  N_{l}\right)  \right\rangle .
\end{equation}
Then%
\begin{equation}
\frac{\left\langle N_{l}F\left(  N_{l}\right)  \right\rangle }{\left\langle
\left(  N_{l}+1\right)  F\left(  N_{l}+1\right)  \right\rangle }=ze^{-x}.
\end{equation}
Repeating this procedure proves Eq. (\ref{NFoNF}).

Now, we can prove Eq. (\ref{Nl}) directly.

We have shown that the statistical distribution (\ref{Nl}) satisfies Eq.
(\ref{FoF}) with $F\left(  N_{l}\right)  =\left(  N_{l}-p_{k_{1}}\right)
\left(  N_{l}-p_{k_{2}}\right)  $. Then, for an arbitrary value of $m$, from
Eq. (\ref{NFoNF}), we can see that the distribution (\ref{Nl}) is a solution
of%
\begin{equation}
\left\langle \left(  N_{l}-p_{k_{1}}\right)  \left(  N_{l}-p_{k_{2}}\right)
N_{l}^{m}\right\rangle =e^{-\beta\left(  \varepsilon_{l}-\mu\right)
}\left\langle \left(  N_{l}+1-p_{k_{1}}\right)  \left(  N_{l}+1-p_{k_{2}%
}\right)  \left(  N_{l}+1\right)  ^{m}\right\rangle ,\text{ \ \ }%
m=0,1,2,\cdots,
\end{equation}
and then is a solution of Eq. (\ref{ExpeqGN}).

This proves the statement that Eq. (\ref{Nl}) is the statistical distribution
corresponding to the operator relation (\ref{adbN}).

This is a general result. For a physical system, we also need to take some
physical conditions into account. We will show that only the first two
nonnegative integer zeroes of $\Lambda\left(  N_{l}\right)  $ contribute to
the statistical distribution of a physical system, though when $p_{k_{1}}$ and
$p_{k_{2}}$ are complex, Eq. (\ref{Nl}) is still a solution of Eq.
(\ref{dmeq0}).

The distribution (\ref{Nl}) is a statistical distribution with both maximum
occupation number $p_{k_{2}}-1$ and minimum occupation number $p_{k_{1}}$.
This is because, by Eq. (\ref{Nl}), when $T\rightarrow0$, the occupation
number of the ground state is $\left\langle N_{0}\right\rangle =p_{k_{2}}-1$.
The minimum value of Eq. (\ref{Nl}) is $p_{k_{1}}$; this tells us that
$p_{k_{1}}$ is the minimum occupation number. In a word, in such a system, a
quantum state can be occupied by at most $p_{k_{2}}-1$ and at least $p_{k_{1}%
}$ particles.

The above result shows that the zeroes of $\Lambda\left(  N_{l}\right)  $
correspond to the restriction on the occupation number of a quantum state.

In a physical system, the minimum occupation number must be zero, i.e.,
$p_{k_{1}}=0$. This requires that for a physical system, the function
$\Lambda\left(  N_{l}\right)  $ must have a zero $N_{l}=0$.

As shown above, $p_{k_{2}}-1$ corresponds to the maximum occupation number. If
we insist on that the maximum occupation number must be a positive integer,
then $p_{k_{2}}$ should be an integer.

For a physical system, if $\Lambda\left(  N_{l}\right)  $ has more than one
zero, only the first two nonnegative integer zeroes are physically meaningful:
the first zero should be $p_{0}=0$, which is the minimum occupation number,
and the second zero gives the maximum occupation number. The other zeroes do
not contribute to the statistical distribution. The reason is as follows.

$p_{k}$ being a zero means that $\Lambda\left(  p_{k}\right)  =0$. Then, from
Eqs. (\ref{badN}) and (\ref{adbN}), we have $b_{l}a_{l}^{\dagger}\left\vert
p_{k}-1\right\rangle =0$ and $a_{l}^{\dagger}b_{l}\left\vert p_{k}%
\right\rangle =0$. By Eq. (\ref{CAactState}), we achieve $\alpha\left(
p_{k}\right)  \beta\left(  p_{k}\right)  =0$. A realistic system must have an
equilibrium state, so $\left\vert \alpha\left(  p_{k}\right)  \right\vert
=\left\vert \beta\left(  p_{k}\right)  \right\vert $. Consequently, we have
$\alpha\left(  p_{k}\right)  =\beta\left(  p_{k}\right)  =0$. This means that
starting from the state with no particle, one cannot achieve the state with
$p_{k}$ particles by repeatedly acting the creation operator.

Now, we can draw our conclusion:

If the relation between the number operator and the creation and annihilation
operators can be expressed as $a^{\dagger}b=\Lambda\left(  N\right)  $, i.e.,
$N$ is a function of quadratic product of the creation and annihilation
operators, then the corresponding statistical distribution is determined by
the first two nonnegative integer zeroes. The first nonnegative integer zero
determines the minimum occupation number, which should be zero for physical
reasons, and the second nonnegative integer zero determines the maximum
occupation number. The corresponding statistical distribution is the Gentile
statistics distribution \cite{Gentile,OursAnn}. Two special cases are that the
maximum occupation number equals $1$, which recovers the Fermi-Dirac
distribution, and that the maximum occupation number tends to $\infty$, which
recovers the Bose-Einstein distribution \cite{OurPhysica,OurPRA}.

It should be emphasized that the above conclusion is based on the demand that
the maximum occupation number (not the average occupation number) is an
integer. If we release this condition, we can obtain other statistical distributions.

\subsection{Examples}

As examples, we first consider the Bose-Einstein, Fermi-Dirac, and Gentile cases.

The above result shows that if the number operator can be written as a
function of the quadratic product of the creation and annihilation operators
$a^{\dagger}b$, i.e., $a^{\dagger}b=\Lambda\left(  N\right)  $, or,
$N=\Lambda^{-1}\left(  a^{\dagger}b\right)  $, then the corresponding
statistical distribution is determined by the zeroes of the function
$\Lambda\left(  N\right)  $ and the distribution must take the form of Eq.
(\ref{Nl}). This means that, if we insist that the maximum occupation number
must be integer, the only physically allowed statistical distributions are
Bose-Einstein, Fermi-Dirac, and Gentile statistics (the distribution with a
non-zero minimum occupation number has no physical meaning).

\textit{(1) The Bose case.} For the Bose case,%
\begin{align}
a^{\dagger}a  &  =\Lambda\left(  N\right)  =N,\\
aa^{\dagger}  &  =\Lambda\left(  N+1\right)  =N+1.
\end{align}
The quantization condition reads%
\begin{equation}
\left[  a,a^{\dagger}\right]  =\Lambda\left(  N+1\right)  -\Lambda\left(
N\right)  =1.
\end{equation}
The only zero of $\Lambda\left(  N\right)  $ is $N=0$. This means that in the
Bose case, the minimum occupation number is $0$, and there is no restriction
on the maximum occupation number. The corresponding statistical distribution
is the Bose-Einstein distribution.

\textit{(2) The Fermi case.} For the Fermi case,%
\begin{align}
a^{\dagger}a  &  =\Lambda\left(  N\right)  =N\left(  2-N\right)  ,\\
aa^{\dagger}  &  =\Lambda\left(  N+1\right)  =\left(  N+1\right)  \left(
1-N\right)  .
\end{align}
The quantization condition reads%
\begin{equation}
\left[  a,a^{\dagger}\right]  =\Lambda\left(  N+1\right)  -\Lambda\left(
N\right)  =1-2N \label{FermiCR}%
\end{equation}
or%
\begin{equation}
\left\{  a,a^{\dagger}\right\}  =\Lambda\left(  N+1\right)  +\Lambda\left(
N\right)  =1+2N-2N^{2}=1.
\end{equation}
Note that for the Fermi case, the value of $N$ can only take $0$ or $1$, so
$1+2N-2N^{2}$ always equals $1$.

The zeroes of $\Lambda\left(  N\right)  $ are $N=0$ and $N=2$. This means that
the maximum occupation number in the Fermi case is $1$. The corresponding
statistical distribution is the Fermi-Dirac distribution.

\textit{(3) The Gentile case.} We can also consider another kind of
intermediate statistics --- Gentile statistics. In Gentile statistics, a
quantum state can be occupied by at most $n$ particles; $n\rightarrow\infty$
and $1$ recover Bose-Einstein statistics \cite{OursAnn,OurDoBose} and
Fermi-Dirac statistics, respectively. For Gentile statistics, two operator
realizations can be constructed as follows:

One is%
\begin{align}
a^{\dagger}a  &  =\Lambda\left(  N\right)  =\frac{\sin\frac{N\pi}{n+1}}%
{\sin\frac{\pi}{n+1}},\\
aa^{\dagger}  &  =\Lambda\left(  N+1\right)  =\frac{\sin\frac{\left(
N+1\right)  \pi}{n+1}}{\sin\frac{\pi}{n+1}}.
\end{align}
The quantization condition reads%
\[
aa^{\dagger}-e^{i\frac{\pi}{n+1}}a^{\dagger}a=e^{-i\frac{N\pi}{n+1}}%
\]
or%
\begin{equation}
aa^{\dagger}-\cos\frac{\pi}{n+1}a^{\dagger}a=\cos\frac{N\pi}{n+1}.
\end{equation}

The other is \cite{OurPhysica,OurPRA}
\begin{align}
a^{\dagger}b  &  =\Lambda\left(  N\right)  =\frac{1-e^{i2\pi\frac{N}{n+1}}%
}{1-e^{i2\pi\frac{1}{n+1}}},\\
ba^{\dagger}  &  =\Lambda\left(  N+1\right)  =\frac{1-e^{i2\pi\frac{N+1}{n+1}%
}}{1-e^{i2\pi\frac{1}{n+1}}}.
\end{align}
Then the quantization condition can be constructed as
\begin{equation}
\left[  b,a^{\dagger}\right]  =e^{i2\pi\frac{N}{n+1}}%
\end{equation}
or%
\begin{equation}
ba^{\dagger}-e^{i2\pi\frac{1}{n+1}}a^{\dagger}b=1.
\end{equation}

The first two zeroes of $\Lambda\left(  N\right)  $ are $N=0$ and $N=n+1$.
This means that the maximum occupation number is $n$ and the corresponding
statistical distribution is the Gentile distribution:%
\begin{equation}
\left\langle N_{l}\right\rangle =\frac{1}{z^{-1}e^{\beta\varepsilon_{l}}%
-1}-\frac{n+1}{z^{-\left(  n+1\right)  }e^{\left(  n+1\right)  \beta
\varepsilon_{l}}-1}. \label{Gd}%
\end{equation}

As a physical realization of Gentile statistics, Ref. \cite{OurJSATA} shows
that the elementary excitation of the Heisenberg magnetic system ---- the
magnon which is the quantized spin waves ---- obeys Gentile statistics with a
maximum occupation number $n=2S$ rather than Bose--Einstein statistics, where
$S$ is the spin quantum number. In the conventional treatment of a magnetic
system, one uses the Holstein--Primakoff realization. The Holstein--Primakoff
realization is a bosonic realization with an additional putting-in-by-hand
restriction on the occupation number. Nevertheless, instead of the bosonic
Holstein--Primakoff realization, Ref. \cite{OurJSATA} constructs a
Gentile-type operator realization, in which there is no additional constraint.
By comparing with the experimental data, in Ref. \cite{OurJSATA}, one can see
that the Gentile realization is more accurate than that of the bosonic
realization. In other words, for a magnetic system, a bosonic realization with
a restriction on the occupation number is indeed an approximation of the
Gentile scheme.

\section{The statistical distributions corresponding to various $q$%
-deformation schemes\label{qdeformation}}

\subsection{The $q$-deformation distributions}

In this section, as the examples of the above general result, we discuss the
statistical distributions for various $q$-deformation schemes.

There are many schemes of $q$-deformation \cite{BDK,BD}, and there are many
discussions on the relation between the operator realization and the
corresponding statistical distribution. However, our result will show that the
statistical distributions given in literature are inaccurate or incomplete.
Our result shows that the $q$-deformation statistical distributions are the
Gentile distribution with a maximum occupation number which is determined by
the deformation parameter $q$ (here we regard Bose-Einstein and Fermi-Dirac
statistics as special cases of Gentile statistics with maximum occupation
numbers $\infty$ and $1$). That is to say, different $q$-deformation schemes
correspond to different kinds of Gentile statistics whose maximum occupation
numbers lie on the deformation parameter $q$.

\textit{(1) }$a^{\dagger}a=\Lambda\left(  N\right)  =\frac{q^{N}-q^{-N}%
}{q-q^{-1}}$\textit{.} In this scheme \cite{Bie,SF},
\begin{equation}
aa^{\dagger}=\Lambda\left(  N+1\right)  =\frac{q^{N+1}-q^{-\left(  N+1\right)
}}{q-q^{-1}}.
\end{equation}
The quantization condition reads%
\begin{equation}
aa^{\dagger}-qa^{\dagger}a=\Lambda\left(  N+1\right)  -q\Lambda\left(
N\right)  =q^{-N}.
\end{equation}

The statistical distribution is determined by the zeroes of $\Lambda\left(
N\right)  $, given by $\left(  q^{N}-q^{-N}\right)  /\left(  q-q^{-1}\right)
=0$.

There are two possible cases: $q=e^{i\frac{k}{2l}2\pi}$ and $q\neq
e^{i\frac{k}{2l}2\pi}$, where $l$ is an arbitrary positive integer, $1\leq
k\leq2l-1$, and $k$ and $2l$ are relatively prime.

For $q=e^{i\frac{k}{2l}2\pi}$, there are many zeroes: $N=0$ and $N$ equals any
integral multiple of $2l$. $N=0$ is the minimum occupation number, and $2l-1$
is the maximum occupation number. This is just Gentile statistics with the
maximum occupation number $2l-1$, i.e., the distribution (\ref{Gd}) with
$n=2l-1$. In this case, the relation between maximum occupation number and $q
$ is%
\begin{equation}
n=i\frac{2\pi k}{\ln q}-1\text{\ with }q=e^{i\frac{k}{2l}2\pi}.
\end{equation}

For $q\neq e^{i\frac{k}{2l}2\pi}$, the only zero of $\Lambda\left(  N\right)
$ is $N=0$. This means that the minimum occupation number is $0$ and there is
no restriction on the maximum occupation number. This is just the
Bose-Einstein distribution, the Gentile distribution with an infinite maximum
occupation number.

\textit{(2) }$a^{\dagger}a=\Lambda\left(  N\right)  =\frac{q^{N}-1}{q-1}%
$\textit{. }In this scheme, the quantization condition is $aa^{\dagger
}-qa^{\dagger}a=1$ \cite{AC} and%
\begin{equation}
\left[  a,a^{\dagger}\right]  =q^{N}.
\end{equation}

Since $\Lambda\left(  N\right)  $ must be real for any value of $N$, $q$
should be a real number. In this scheme, $0<q<1$. Therefore, the only zero is
$N=0$ and the corresponding statistical distribution is the Bose-Einstein distribution.

\textit{(3) }$a^{\dagger}a=\Lambda\left(  N\right)  =\frac{q^{N}-p^{-N}%
}{q-p^{-1}}$\textit{. }In this scheme, the quantization condition is
$aa^{\dagger}-qa^{\dagger}a=p^{-N}$ and $aa^{\dagger}-p^{-1}a^{\dagger}a=q^{N}
$ \cite{CJ1991,BJM}. Then we have%
\begin{equation}
\left[  a,a^{\dagger}\right]  =\frac{q^{N}\left(  q-1\right)  +p^{-N}\left(
1-p^{-1}\right)  }{q-p^{-1}}.
\end{equation}

When
\begin{equation}
q=\left\vert q\right\vert e^{i\frac{k}{2l}2\pi}\text{ \ and \ }p=\left\vert
q\right\vert ^{-1}e^{i\frac{k}{2l}2\pi},
\end{equation}
where $l$ is a positive integer and $k$ and $2l$ are relatively prime, the two
minimum zeroes of $\Lambda\left(  N\right)  $ are
\begin{equation}
N=0\text{ and }N=2l.
\end{equation}
Then the corresponding distribution is the Gentile distribution with the
maximum occupation number $2l-1$, the distribution (\ref{Gd}) with $n=2l-1$.

When $q$ and $p$ take other values, there is only one zero $N=0$, and the
distribution is the Bose-Einstein distribution.

That is to say, for a given $q$ and $p$, this scheme corresponds to Gentile
statistics with a maximum occupation number determined by the deformation
parameters $q$ and $p$.

\textit{(4) }$a^{\dagger}a=\Lambda\left(  N\right)  =N\left(  p+1-N\right)  $.
In the parafermionic scheme, the matrix element of $a^{\dagger}$ and $a$ is
$a_{N,N+1}=a_{N+1,N}^{\dagger}=\sqrt{\left(  N+1\right)  \left(  p-N\right)
}$ \cite{OK}. The quantization condition then reads%
\begin{equation}
\left[  a,a^{\dagger}\right]  =p-2N.
\end{equation}

When $p$ is a positive integer, $\Lambda\left(  N\right)  $ has two zeroes%
\begin{equation}
N=0\text{ and }N=p+1\text{.}%
\end{equation}
The corresponding statistical distribution is the Gentile distribution with
the maximum occupation number $p$, the distribution (\ref{Gd}) with $n=p$.

Such a result can be directly seen from the operator realization of
parafermionic quantization.

For $p=1$, the operator realization is just the fermionic case: $\left\{
a,a^{\dagger}\right\}  =1$ \cite{OK}. This leads to $a^{2}=0$, i.e., the
maximum occupation number is $1$. For $p=2$, the operator realization is:
$a^{3}=0$, $aa^{\dagger}a=2a$, and $aaa^{\dagger}+a^{\dagger}aa=2a$ \cite{OK}.
$a^{3}=0$ means that the maximum occupation number is $2$. In other words, one
can recognizes from the operator realization of the parafermionic quantization
that such a $q$-deformation scheme corresponds to Gentile statistics.

When $p$ is not a positive integer, $\Lambda\left(  N\right)  $ has only one
zero $N=0$, the distribution is the Bose-Einstein distribution.

\textit{(5) }$a^{\dagger}a=\Lambda\left(  N\right)  =\frac{\sinh\left(  \tau
N\right)  \sinh\left[  \tau\left(  p+1-N\right)  \right]  }{\sinh^{2}\left(
\tau\right)  }$. In the $q$-deformed para-Fermi quantization scheme
\cite{OKK,FV}, $a^{\dagger}a=\left[  -\left(  N-1-p\right)  \right]  \left[
N\right]  $, where $\left[  x\right]  =\frac{q^{x}-q^{-x}}{q-q^{-1}}$. The
commutation relation, by the substitution $\tau=\ln q$, then reads%
\begin{equation}
\left[  a,a^{\dagger}\right]  =\frac{\cosh\left[  \tau\left(  1+p-2N\right)
\right]  -\cosh\left[  \tau\left(  1-p+2N\right)  \right]  }{2\sinh^{2}\left(
\tau\right)  }.
\end{equation}
and $\Lambda\left(  N\right)  $ can be rewritten as $\frac{\sinh\left(  \tau
N\right)  \sinh\left[  \tau\left(  p+1-N\right)  \right]  }{\sinh^{2}\left(
\tau\right)  }$.

Like that in the above case, when $p$ is a positive integer, $\Lambda\left(
N\right)  $ has two zeroes%
\begin{equation}
N=0\text{ and }N=p+1.
\end{equation}
The distribution is the Gentile distribution with the maximum occupation
number $p$. When $p$ takes another value, $\Lambda\left(  N\right)  $ has only
one zero, the distribution is the Bose-Einstein distribution.

\textit{(6) }$a^{\dagger}a=\Lambda\left(  N\right)  =N\cos^{2}\frac{N\pi}%
{2}+\left(  N+p-1\right)  \sin^{2}\frac{N\pi}{2}$. In the parabosonic scheme
\cite{OK}, for $p=1$, the operator realization is $\left[  a,a^{\dagger
}\right]  =1$ and this is just the Bose case; for $p=2$, the operator
realization is $aaa^{\dagger}-a^{\dagger}aa=2a$, etc. The matrix elements of
$a^{\dagger}$ and $a$ are $a_{N,N+1}=a_{N+1,N}^{\dagger}=\sqrt{N+p}$
($N=even$) and $a_{N,N+1}=a_{N+1,N}^{\dagger}=\sqrt{1+N}$ ($N=$ $odd$). Then
$a^{\dagger}a=N+p-1$ for $N=$ $odd$ and $a^{\dagger}a=N$ for $N=even$. Such
results can be equally rewritten in a compact form: $a_{N,N+1}=a_{N+1,N}%
^{\dagger}=\cos^{2}\frac{N\pi}{2}\sqrt{N+p}+\sin^{2}\frac{N\pi}{2}\sqrt{1+N}$
and $a^{\dagger}a=N\cos^{2}\frac{N\pi}{2}+\left(  N+p-1\right)  \sin^{2}%
\frac{N\pi}{2}$. Then the quantization condition reads%
\begin{equation}
\left[  a,a^{\dagger}\right]  =2\left(  1-p\right)  \sin^{2}\frac{N\pi}{2}+p.
\end{equation}
In this case, $\left\{  a,a^{\dagger}\right\}  =2N-p$.

When $p=1-\frac{l}{\sin^{2}\left(  l\pi/2\right)  }$, where $l>1$ is an
integer, $\Lambda\left(  N\right)  $ has two zeroes%
\begin{equation}
N=0\text{ and }N=l,
\end{equation}
the corresponding distribution is the Gentile distribution with the maximum
occupation number $l-1$, the distribution (\ref{Gd}) with $n=l-1$.

When $p$ takes another value, $\Lambda\left(  N\right)  $ has only one zero
$N=0$, the distribution is the Bose-Einstein distribution.

\textit{(7) }$a^{\dagger}a=\Lambda\left(  N\right)  =\frac{\sinh\left(  \tau
N\right)  }{\sinh\tau}\frac{\cosh\left[  \tau\left(  N+2N_{0}-1\right)
\right]  }{\cosh\tau}\cos^{2}\frac{N\pi}{2}+\frac{\sinh\left[  \tau\left(
N+2N_{0}-1\right)  \right]  }{\sinh\tau}\frac{\cosh\left(  \tau N\right)
}{\cosh\tau}\sin^{2}\frac{N\pi}{2}$. In the $q$-deformed Wigner quantization
scheme \cite{OKK,FV}, the matrix elements of $a^{\dagger}$ and $a$ are
$a_{N,N+1}=a_{N+1,N}^{\dagger}=\sqrt{\left[  N+2N_{0}\right]  \left\{
N+1\right\}  }$ ($N=even$) and $a_{N,N+1}=a_{N+1,N}^{\dagger}=\sqrt{\left\{
N+2N_{0}\right\}  \left[  N+1\right]  }$ ($N=odd$), where $\left\{  x\right\}
=\frac{q^{x}+q^{-x}}{q+q^{-1}}$. Then we have $a^{\dagger}a=\left[
N-1+2N_{0}\right]  \left\{  N\right\}  $ ($N=odd$) and $a^{\dagger}a=\left\{
N-1+2N_{0}\right\}  \left[  N\right]  $ ($N=even$). Substituting $\tau=\ln q$
and rewriting the above results in a compact form, we have $a_{N,N+1}%
=a_{N+1,N}^{\dagger}=\sqrt{\left\{  N+2N_{0}\right\}  \left[  N+1\right]
}\sin^{2}\frac{N\pi}{2}+$ $\sqrt{\left[  N+2N_{0}\right]  \left\{
N+1\right\}  }$ $\cos^{2}\frac{N\pi}{2}$ and $a^{\dagger}a=\frac{\sinh\left(
\tau N\right)  }{\sinh\tau}\frac{\cosh\left[  \tau\left(  N+2N_{0}-1\right)
\right]  }{\cosh\tau}\cos^{2}\frac{N\pi}{2}+\frac{\sinh\left[  \tau\left(
N+2N_{0}-1\right)  \right]  }{\sinh\tau}\frac{\cosh\left(  \tau N\right)
}{\cosh\tau}\sin^{2}\frac{N\pi}{2}$. The commutation relation then reads
\begin{equation}
\left[  a,a^{\dagger}\right]  =\frac{\cosh\left[  2\tau\left(  N+N_{0}\right)
\right]  }{\cosh\tau}-2\cos\left(  N\pi\right)  \frac{\sinh\left[  \tau\left(
1-2N_{0}\right)  \right]  }{\sinh\left(  2\tau\right)  }.
\end{equation}
In this case, $\left\{  a,a^{\dagger}\right\}  =\frac{\sinh\left[
2\tau\left(  N+N_{0}\right)  \right]  }{\sinh\tau}$.

When $N_{0}>0$, $\Lambda\left(  N\right)  $ has only one zero $N=0$, the
distribution is the Bose-Einstein distribution.

\textit{(8) }$a^{\dagger}a=\Lambda\left(  N\right)  =\sin^{2}\frac{N\pi}{2} $.
In this scheme \cite{JB},
\begin{equation}
\left[  a,a^{\dagger}\right]  =\left(  -1\right)  ^{N}. \label{scheme8}%
\end{equation}
In this case, $\left\{  a,a^{\dagger}\right\}  =1$. Note that this scheme is
just the scheme \textit{(2)} with $q=-1$, i.e., $a^{\dagger}a=\left.
\frac{q^{N}-1}{q-1}\right\vert _{q=-1}$.

The two minimum zeroes of $\Lambda\left(  N\right)  $ are $N=0$ and $2$. This
means that the maximum occupation number is $1$, and then the corresponding
statistical distribution is just the Fermi-Dirac distribution. This is
because, from Eq. (\ref{FermiCR}), in the Fermi case we have $\left[
a,a^{\dagger}\right]  =1-2N$. Moreover, in the Fermi case, the value of $N$
can take on only $0$ and $1$, while in Eq. (\ref{scheme8}), the values of
$\left(  -1\right)  ^{N}$ are only $+1$ and $-1$.

\textit{(9) }$a^{\dagger}a=\Lambda\left(  N\right)  =q^{N-1}\sin^{2}\frac
{N\pi}{2}$\textit{.} In this scheme \cite{Hay},%
\begin{equation}
\left[  a,a^{\dagger}\right]  =\left(  q^{N}-q^{N-1}\right)  +\left(
-1\right)  ^{N}\left(  q^{N}+q^{N-1}\right)  .
\end{equation}

Like the above case, the two minimum zeroes of $\Lambda\left(  N\right)  $ are
$N=0$ and $2$, and the distribution is the Fermi-Dirac distribution.

It can be seen that the statistical distribution of case (8) and case (9) are
the same. This is because the type of statistical distribution is only
determined by the zero of $\Lambda\left(  N\right)  $. So long as the zeroes
of $\Lambda\left(  N\right)  $\ are the same, the statistical distributions
are the same.

\textit{(10) }$a^{\dagger}a=\Lambda\left(  N\right)  =\frac{1-\left(
-q\right)  ^{N}}{1+q}$. In this scheme \cite{VPJ}, the operator realization is
$aa^{\dagger}+qa^{\dagger}a=1$. We then have%
\begin{equation}
\left[  a,a^{\dagger}\right]  =\left(  -q\right)  ^{N}.
\end{equation}

When $q=1$, the two minimum zeroes of $\Lambda\left(  N\right)  $ are $N=0$
and $2$, and the distribution is the Fermi-Dirac distribution.

When $q\neq1$, there is only one zero $N=0$; the distribution is the
Bose-Einstein distribution.

\textit{(11) }$a^{\dagger}a=\Lambda\left(  N\right)  =N^{n}$. In this scheme
\cite{Das}, the operator realization is $\left(  aa^{\dagger}\right)
^{n}-\left(  a^{\dagger}a\right)  ^{n}=1$. We then have%
\begin{equation}
\left[  a,a^{\dagger}\right]  =\left(  N+1\right)  ^{n}-N^{n}.
\end{equation}
For any value of $n$, there is only one zero; the distribution is the
Bose-Einstein distribution.

\subsection{Comment on the literature result of the $q$-deformation
statistical distribution \label{comment}}

In literature, there are many discussions on the statistical distributions for
various $q$-deformation schemes, especially the scheme $aa^{\dagger
}-qa^{\dagger}a=q^{-N}$ with $a^{\dagger}a=\Lambda\left(  N\right)  =\left(
q^{N}-q^{-N}\right)  /\left(  q-q^{-1}\right)  $. However, all these results
on the $q$-deformation statistical distribution are inaccurate or incomplete.

When calculating the statistical distribution from a given operator relation,
one may encounter a problem of the calculation of $\left\langle f\left(
N\right)  \right\rangle $, the average of a function of the number operator
$N$. In some literature, the authors use an approximation%
\begin{equation}
\left\langle f\left(  N\right)  \right\rangle =f\left(  \left\langle
N\right\rangle \right)  , \label{imapp}%
\end{equation}
i.e., approximately replace the average of the function of the number operator
$N$ by the function of the average of the number operator $\left\langle
N\right\rangle $. Concretely, for example, in Ref. \cite{TRM}, the authors use
the approximation $q^{\left\langle N\right\rangle }\simeq\left\langle
q^{N}\right\rangle $. This gives $\left[  \left\langle N\right\rangle \right]
\simeq\left\langle \left[  N\right]  \right\rangle =\left\langle a^{\dagger
}a\right\rangle $. This result has been used as the basis of some other works
by many authors, e.g., \cite{Alg,SCC,SCC2,SCC3,CSC,SS,LS}. Moreover, such a
treatment also appears in other literature \cite{AA,AA2,KLQ,LSS,Duc,LY}. In
fact, some authors had noticed that the distribution obtained by such a way is
not properly the distribution of the average occupation number \cite{SS}. The
statistical distribution they obtained is of course not the Gentile
distribution. Nevertheless, as analyzed above, the statistical distributions
corresponding to such $q$-deformation schemes are Gentile distributions with
various maximum occupation numbers determined by the deformation parameter $q$.

In literature, there is also an alternative way for considering the
$q$-deformation statistical distribution \cite{AGI,AG,AGI2,SDA}: replacing the
number operator $N$ by $a^{\dagger}a$. In other words, in this treatment it is
$\Lambda\left(  N\right)  $, rather than $N$, that plays the role of the
number operator; for example, in the $q$-deformation scheme $aa^{\dagger
}-qa^{\dagger}a=q^{-N}$, $\left[  N\right]  =\left(  q^{N}-q^{-N}\right)
/\left(  q-q^{-1}\right)  $ plays the role of number operator rather than $N$
itself. The reason why such a replacement is adopted is that, as the authors
state, if still using $N$ as the number operator, the corresponding
statistical distribution is a nondeformed Bose-Einstein distribution
\cite{DK,KMD,CFM}. This statement is incomplete. According to our above
result, taking the scheme $aa^{\dagger}-qa^{\dagger}a=q^{-N}$ as an example,
we can see that even if starting from $N$ rather than $a^{\dagger}a$ to
construct the statistical distribution, there is still a deformed distribution
when $q=e^{i\frac{k}{2l}2\pi}$: the Gentile distribution with the maximum
occupation number $2l-1$.

\section{Conclusions and outlook\label{conclusion}}

In this paper, we give a general discussion on the calculation of the
statistical distribution from a given operator relation. Our result shows that
as long as the relation between the number operator and creation and
annihilation operators can be expressed as $a^{\dagger}b=\Lambda\left(
N\right)  $, i.e., $N$ is a function of quadratic product of the creation and
annihilation operators, the corresponding statistical distribution is
determined by the first two nonnegative integer zeroes of the function
$\Lambda\left(  N\right)  $. The statistical distribution cannot be anything
but the Gentile distribution, including the two limit cases of the Gentile
distribution ---- Bose-Einstein and Fermi-Dirac distributions.

On the basis of the general result of the relation between the operator
relation and the statistical distribution, we systematically discuss the
statistical distributions of various $q$-deformation schemes. We point out
that the statistical distributions corresponding to all $q$-deformation
schemes are Gentile distributions with various maximum occupation numbers
which are determined by the deformation parameter $q$.

Our result shows that for the statistical distribution of the $q$-deformation
scheme, there are two inaccurate results in the literature: (1) replacing
$\left\langle f\left(  N\right)  \right\rangle $ by $f\left(  \left\langle
N\right\rangle \right)  $ in the calculation of the statistical distribution,
and (2) considering that the statistical average $\left\langle N_{l}%
\right\rangle $ gives only the Bose-Einstein distribution.

As an outlook, we would like to suggest several topics worth to be considered
in further studies.

The operator relations considered in the present paper are only of the form
$a^{\dagger}b=\Lambda\left(  N\right)  $ or $N=\Lambda^{-1}\left(  a^{\dagger
}b\right)  $. Nevertheless, our result also implies that the operator
relations corresponding to many intermediate statistical distributions are not
the Gentile distribution, e.g., the Haldane-Wu statistics, and cannot be
expressed as $a^{\dagger}b=\Lambda\left(  N\right)  $. Therefore, we also need
to consider the intermediate statistics corresponding to more general operator relations.

The statistical distributions corresponding to some $q$-deformation schemes,
as shown in the present paper, are Gentile distributions with different
maximum occupation numbers. In further studies, we can consider more general
cases. Concretely, Gentile statistics is a special case of generalized
statistics introduced in Ref. \cite{OurGS}. In the generalized statistics, the
maximum occupation numbers of different quantum states can take on different
values. Starting from the generalized statistics, one can construct a system
with both bosonic states and fermionic states. For example, Ref. \cite{OurGS}
constructs an exactly solvable phase transition model which shows that a
system with one bosonic ground state and fermionic excited states can display
Bose-Einstein-condensation type phase transition. By generalizing the result
of the present paper, we can build a set of operator relations for the
generalized statistics.

Resent researches show that there are some physical systems, such as a
Heisenberg magnetic system, obey Gentile statistics rather than Bose-Einstein
or Fermi-Dirac statistics \cite{OurJSATA}. As shown in the present paper, many
$q$-deformation operator schemes correspond to Gentile statistics with a
maximum occupation number determined by $q$. This inspires us to apply various
$q$-deformation operator schemes to such physical systems.

The idea of $q$-analog has wide-ranging applications in many areas, e.g., in
statistical mechanics \cite{Tsa1,Tsa2} and in probability theory
\cite{qGauss,LZDX}. In physics, an important $q$-analog is the quantum
algebra. The applications of quantum algebras cover many physical fields, such
as nuclear physics \cite{BD,SBG}, gravity \cite{KC,BR}, noncommutative
space-time \cite{FGB}, supersymmetric Yang-Mills \cite{GH}, string theory
\cite{HHM}, quantum entanglement \cite{SPK}, and statistical physics
\cite{SS2008,GKM,AS}. The bridge between operator relation and statistical
distribution allows us to study such systems more deeply.

\vskip0.5cm

We are very indebted to Dr G Zeitrauman for his encouragement. This work is
supported in part by NSF of China under Grant No. 11075115.


\begin{thebibliography}{99}                                                                                               %


\bibitem {Gentile}G. Gentile, Nuovo Cimento 17 (1940) 493; A. Khare,
Fractional Statistics and Quantum Theory, World Scientific, Singapore, 1997.

\bibitem {gPauli}F.D.M. Haldane, Phys. Rev. Lett. 67 (1991) 937.

\bibitem {Wu}Y.-S. Wu, Phys. Rev. Lett. 73 (1994) 922.

\bibitem {Wilczek}F. Wilczek, Phys. Rev. Lett. 48 (1982) 1144; F. Wilczek,
Phys. Rev. Lett. 49 (1982) 957.

\bibitem {Kitaev}A. Kitaev, Ann. Phys. (N.Y.) 321 (2006) 2; Z.N.C. Ha, Phys.
Rev. Lett. 73 (1994) 1574; B. Sutherland, Phys. Rev. B 56 (1997) 4422; M.
Wadati, J. Phys. Soc. Jpn 64 (1995) 1552.

\bibitem {FermiSuperfluid}B.DeMarco, D.S. Jin, Science 285 (1999) 1703; A.G.
Truscott, K.E. Strecker, W.I. McAlexander, G.B. Partridge, R.G. Hulet, Science
291 (2001) 2570; F. Schreck, L. Khaykovich, K.L. Corwin, G. Ferrari, T.
Bourdel, J. Cubizolles, C. Salomon, Phys. Rev. Lett. 87 (2001) 080403; Y.
Ohashi, A. Griffin, Phys. Rev. Lett. 89 (2002) 130402.

\bibitem {exciton}L.V. Butov, A.C. Gossard, D.S. Chemla, Nature 418 (2002) 751.

\bibitem {OurJSATA}W.-S. Dai, M. Xie, J. Stat. Mech. (2009) P04021.

\bibitem {Guan}X.W. Guan, M.T. Batchelor, C. Lee, M. Bortz, Phys. Rev. B 76
(2007) 085120.

\bibitem {CompositeParticle}E. Hanamura, H. Haug, Phys. Rep. 33 (1977) 209.

\bibitem {OursAnn}W.-S. Dai, M. Xie, Ann. Phys. (N.Y.) 309 (2004) 295.

\bibitem {OurDoBose}W.-S. Dai, M. Xie, Phys. Lett. A 373 (2009) 1524.

\bibitem {Drinfeld}V.G. Drinfeld, in: A.M. Gleason (Ed.), Proceedings of the
International Congress of Mathematicians, American Mathematical Society,
Providence, RI, 1986.

\bibitem {Jimbo}M. Jimbo, Lett. Math. Phys. 11 (1986) 247.

\bibitem {AC}M. Arik, D.D. Coon, J. Math. Phys. 17 (1976) 524.

\bibitem {Bie}L.C. Biedenharn, J. Phys. A 22 (1989) L873; A.J. Macfarlane, J.
Phys. A 22 (1989) 4581;

\bibitem {OK}Y. Ohnuki, S. Kamefuchi, Quantum Field Theory and Parastatistics,
Springer Verlag, Berlin, 1982.

\bibitem {OKK}K. Odaka, T. Kishi, S. Kamefuchi, J. Phys. A 24 (1991) L591.

\bibitem {Hay}T. Hayashi, Commun. Math. Phys. 127 (1990) 129; M. Chaichian, P.
Kulish, Phys. Lett. B 234 (1990) 72; L. Frappat, P. Sorba, A. Sciarrino, J.
Phys. A 24 (1991) L179; D. Gangopadhyay, Acta Physica Polonica B 22 (1991)
819; A. Sciarrino, J. Phys. A 25 (1992) L219; R. Parthasarathy, K.S.
Viswanathan, J. Phys. A 24 (1991) 613.

\bibitem {CJ1991}R. Chakrabarti and R. Jagannathan, J. Phys. A 24 (1991) L711.

\bibitem {Das}C. Daskaloyannis, J. Phys. A 24 (1991) L789.

\bibitem {CCN}W.-S. Chung, K.-S. Chung, S.-T. Nam, Phys. Lett. A 183 (1993) 363.

\bibitem {MMP}S. Meljanac, M. Milekovi\'{c}, S. Pallua, Phys. Lett. B 328
(1994) 55; preprint: arXiv hep-th/9404039.

\bibitem {OS}C.H. Oh, K. Singh, J. Phys. A 27 (1994) 5907; arXiv:hep-th/9407142.

\bibitem {Bur}I.M. Burban, J. Phys. A: Math. Theor. 42 (2009) 065201;
preprint: arXiv 1110.1025.

\bibitem {KD}P.P. Kulish, E.V. Damaskinsky, J. Phys. A 23 (1990) L415.

\bibitem {UA}H. Ui, N. Aizawa, Mod. Phys. Lett. A 05 (1990) 237.

\bibitem {SF}C.P. Sun, H.C. Fu, J. Phys. A 22 (1989) L983.

\bibitem {GX}M.-L. Ge, K. Xue, J. Phys. A 24 (1991) L895.

\bibitem {KB}A. Kundu, B. Basu-Mallick, J. Phys. A 27 (1994) 3091.

\bibitem {BH}\v{C}. Burd\'{\i}t, P Hellinger, J. Phys. A 25 (1992) L1023.

\bibitem {CJ}R. Chakrabarti, R. Jagannathan, J. Phys. A 27 (1994) 2023.

\bibitem {BBC}M. Bednar, C. Burdik, M. Couture, L. Hlavaty, J. Phys. A 25
(1992) L341.

\bibitem {SWZ}W.B. Schmidke, J. Wess, B. Zumino, Z. Phys. C 52 (1991) 471.

\bibitem {OSW}O. Ogievetsky, W.B. Schmidke, J. Wess, B. Zumino, Commun. Math.
Phys. 150 (1992) 495.

\bibitem {Ane}B. Aneva, Phys. Lett. B 340 (1994) 155; preprint: 0804.1529

\bibitem {TRM}J.A. Tuszynski, J.L. Rubin, J. Meyer, M. Kibler, Phys. Lett. A
175 (1993) 173.

\bibitem {Alg}A. Algin, J. Stat. Mech. (2008) P10009.

\bibitem {SCC}G. Su, J. Chen, L. Chen, J. Phys. A 36 (2003) 10141.

\bibitem {SCC2}G. Su, S. Cai, J. Chen, J. Phys. A 41 (2008) 045007.

\bibitem {SCC3}Y. Shu, J. Chen, L. Chen, Phys. Lett. A 292 (2002) 309.

\bibitem {CSC}S. Cai, G. Su, J. Chen, J. Phys. A 40 (2007) 11245.

\bibitem {SS}A.M. Scarfone, P.N. Swamy, J. Stat. Mech. (2009) P02055.

\bibitem {LS}A. Lavagno, P.N. Swamy, Phys. Rev. E 65 (2002) 036101.

\bibitem {AA}A. Algin, E. Arslan, Phys. Lett. A 372 (2008) 2767.

\bibitem {AA2}A. Algin, E. Arslan, J. Phys. A 41 (2008) 365006.

\bibitem {KLQ}G. Kaniadakis, A. Lavagno, P. Quarati, Phys. Lett. A 227 (1997) 227.

\bibitem {LSS}A. Lavagno, A.M. Scarfone, P.N. Swamy, Rep. Math. Phys. 55
(2005) 423.

\bibitem {Duc}D.V. Duc, Generalised $q$-deformed oscillators and their
statistics, preprint: arXiv hep-th/9410232

\bibitem {LY}C.R. Lee, J.-P. Yu, Phys. Lett. A 150 (1990) 63.

\bibitem {DK}M. Daoud, M. Kibler, Phys. Lett. A 206 (1995) 13.

\bibitem {KMD}M.R. Kibler, J. Meyer, M. Daoud, On $qp$-deformations in
statistical mechanics of bosons in D dimensions, preprint: arXiv cond-mat/9611042.

\bibitem {CFM}M. Chaichian, R.G. Felipe, C Montonen, J. Phys. A 26 (1993) 4017.

\bibitem {AGI}D.V. Anchishkin, A.M. Gavrilik, N.Z. Iorgov, Mod. Phys. Lett. A
15 (2000) 1637.

\bibitem {AG}L.V. Adamska, A.M. Gavrilik, J. Phys. A 37 (2004) 4787.

\bibitem {AGI2}D.V. Anchishkin, A.M. Gavrilik, N.Z. Iorgov, Eur. Phys. J. A 7
(2000) 229.

\bibitem {SDA}H.-S. Song, S.-X. Ding, I. An, J. Phys. A 26 (1993) 5197.

\bibitem {cal}C. R. Lee and J.P. Yu, Phys. Lett. A150,(1990) 63; C.R. Lee and
J.P. Yu, Phys. Lett. A 164 (1992) 164; J.A. Tuszy\'{n}ski, J.L. Rubin, J.
Meyer, M. Kibler, Phys. Lett. A 175 (1993) 173.

\bibitem {OurPhysica}W.-S. Dai, M. Xie, Physica A 331 (2004) 497.

\bibitem {OurPRA}Y. Shen, W.-S. Dai, M. Xie, Phys. Rev. A 75 (2007) 042111.

\bibitem {BDK}D. Bonatsos, C. Daskaloyannis, P. Kolokotronis, Generalized
deformed oscillators and algebras, preprint: arXiv hep-th/9512083.

\bibitem {BD}D. Bonatsos, C. Daskaloyannis, Prog. Part. Nucl. Phys. 43 (1999)
537; preprint: arXiv nucl-th/9909003.

\bibitem {BJM}A. Jannussis, G. Brodimas, R. Mignani, J. Phys. A 24 (1991) L775;

\bibitem {FV}R. Floreanini, L. Vinet, J. Phys. A 23 (1990) L1019;

\bibitem {JB}A. Jannussis, G. Brodimas, D. Sourlas, V. Zisis, Lett. Nuovo
Cimento 30 (1981) 123.

\bibitem {VPJ}K.S. Viswanathan, R. Parthasarathy, R. Jagannathan, J. Phys. A
25 (1992) L335.

\bibitem {OurGS}W.-S. Dai, M. Xie, J. Stat. Mech. (2009) P07034

\bibitem {Tsa1}C. Tsallis, J. Stat. Phys. 52 (1988) 479.

\bibitem {Tsa2}C. Tsallis, Introduction to Non-extensive Statistical
Mechanics, Springer, 2009.

\bibitem {qGauss}C. Tsallis, A. Rapisarda, A. Pluchino, E.P. Borges, Physica A
381 (2007) 143; A. Rodr\'{\i}guez, V. Schw\"{a}mmle, C. Tsallis, J. of Stat.
Mech. (2008) P09006. P. Douglas, S. Bergamini, F. Renzoni. Phys. Rev. Lett. 96
(2006) 110601.

\bibitem {LZDX}T. Liu, P. Zhang, W.-S. Dai, and Mi Xie, Physica A 391 (2012) 5411.

\bibitem {SBG}K.D. Sviratcheva, C. Bahri, A.I. Georgieva, J.P. Draayer, Phys.
Rev. Lett. 93 (2004) 152501; \ \ \ \ \ \ \ \ preprint: arXiv nucl-th 0703068

\bibitem {KC}I. Khavkine, J.D. Christensen, Class. Quantum Grav. 24 (2007)
3271;\ preprint: arXiv 0704.0278.

\bibitem {BR}E. Bianchi, C. Rovelli, Phys. Rev. D 84 (2011) 027502;\ preprint:
arXiv 1105.1898.

\bibitem {FGB}A. Fring, J. Phys. A 43 (2010) 425202; preprint: arXiv 1006.2065.

\bibitem {GH}C. G\'{o}mez and R. Hern\'{a}ndez, JHEP 0703 (2007) 108;
preprint: arXiv hep-th/0701200.

\bibitem {HHM}B. Hoare, T.J. Hollowood, J.L. Miramontes, JHEP 1210 (2012)
76;\ preprint: arXiv 1206.0010.

\bibitem {SPK}R.A. Santos, F.N.C. Paraan, V.E. Korepin, A. Kl\"{u}mper, EPL 98
(2012) 37005;\ preprint: arXiv 1112.0517.

\bibitem {SS2008}A.M. Scarfone, P.N. Swamy, J. Phys. A 41 (2008) 275211;
preprint: arXiv 0807.2510.

\bibitem {GKM}A.M. Gavrilik, I.I. Kachurik, Yu.A. Mishchenko, J. Phys. A 44
(2011) 475303;\ preprint: arXiv 1107.5704.

\bibitem {AS}A. Algin, M. Senay, Phys. Rev. E 85 (2012) 041123; preprint:
arXiv 1205.0926.
\end{thebibliography}
\end{document}